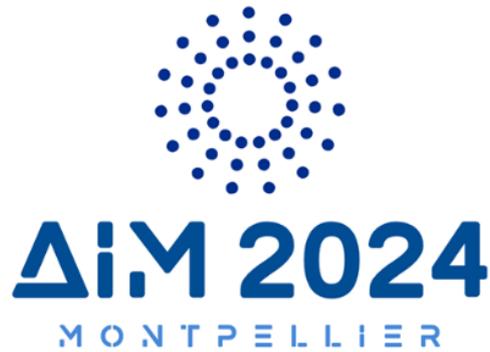

**29ᵉ Conférence de l'Association Information et Management**
*27-29 mai 2024 à Montpellier - La Grande-Motte*

# Révéler la légitimité dans le contexte des événements inattendus : Une enquête sur les sociétés de conseil en systèmes d'information et les organisations internationales grâce à une analyse de Topic Modeling.

Oussama Abidi, Aix Marseille Univ, Université de Toulon, CERGAM, Aix-en-Provence, France, oussama.abidi@etu.univ-amu.fr

**Résumé**

Dans un marché dynamique et moderne, la récurrence d'événements inattendus nécessite des réponses proactives de la part des parties prenantes du système d'information (SI). Chaque acteur du SI s'efforce de légitimer ses actions et de communiquer sa stratégie. Cette étude explore la légitimation en SI, en mettant l'accent sur la communication de deux acteurs clés : les sociétés de conseil en SI et les organisations internationales, dans le contexte d'événements inattendus. Pour atteindre cet objectif, nous avons examiné une gamme diversifiée de documents publiées par les deux parties prenantes. En utilisant la méthodologie de Topic Modeling, nous avons analysé ces documents pour extraire des informations sur leurs méthodes de légitimation. À travers cette recherche, nous visons à contribuer à la littérature sur la communication en offrant une exploration de deux acteurs clés du SI en réponse aux défis posés par des événements inattendus.

**Mot clés**

Communication ; sociétés de conseil en SI ; organisations internationales ; système d'information (SI) ; discours de légitimation.

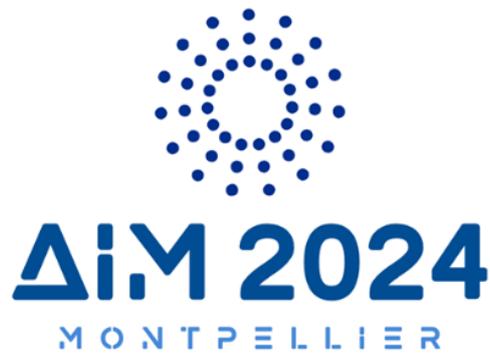

**29ᵉ Conférence de l'Association Information et Management**
*27-29 mai 2024 à Montpellier - La Grande-Motte*

# Unveiling Legitimacy in the unexpected events context : An Inquiry into Information System Consultancy companies and international organizations through Topic Modeling Analysis.

Oussama Abidi, Aix Marseille Univ, Université de Toulon, CERGAM, Aix-en-Provence, France, oussama.abidi@etu.univ-amu.fr

**Abstract**

In an increasingly dynamic and modern market, the recurrence of unexpected events necessitates proactive responses from information system (IS) stakeholders. Each IS actor strives to legitimize its actions and communicate its strategy. This study delves into the realm of IS legitimation, focusing on the communication of two key stakeholders: IS consultancy companies and international organizations, particularly in the context of unexpected events. To achieve this objective, we examined a diverse array of publications released by both actors. Employing a topic modeling methodology, we analyzed these documents to extract valuable insights regarding their methods of legitimation. Through this research, we aim to contribute to the legitimation discourse literature by offering an exploration of two key IS stakeholders responding to the challenges posed by unexpected events.

# 1  Introduction

In today's rapidly evolving and unpredicted market, the role of IS is undeniable. As businesses and institutions increasingly rely on complex IS, the impact of unexpected events, from security breaches to technological malfunction and delays, has never been more profound (Coulon et al., 2023). These events necessitate well-informed responses from diverse stakeholders within the IS spectrum, including IS users, executives and managers, IS professionals, Vendors and suppliers, and regulators and compliance bodies (Hovelja et al., 2013; Pouloudi & Whitley, 1997).

In this paper we delve into the responses to unexpected events from an organizational perspective. Drawing from the insights provided by Hovelja et al. (2013) and Pouloudi & Whitley (1997), we focus on two key stakeholders for in-depth analysis: IS consultancy companies, presenting the vendors and suppliers category, and international organizations, embodying the regulators and compliance bodies category, when confronted with unexpected events. by examining the distinct approaches taken by these stakeholders, we aim to shed light on the nuanced strategies employed in response to unexpected events.

IS consultancy companies invest considerable effort in addressing challenges posed by complex IS and unexpected events, providing valuable insights, field feedback, and innovative services to overcome challenges (Stockhinger & Teubner, 2018). Their active engagement positions them to gather firsthand experiences and innovative ideas (Johnson et al., 2008). Additionally, their exposure to diverse contexts and organizations enables knowledge synthesis (Jarzabkowski & Spee, 2009). Finally, IS consultancy companies' documentation serves as a rich corpus for researchers to analyze (Stockhinger & Teubner, 2018).

Also, the international organizations, such as the United Nations and the World Bank, play significant roles in shaping efforts to address the challenges posed by complex IS and unexpected events. Their funding, regulations, and evaluation mechanisms strongly influence IS research (Guy, 2020). For instance, international policies and regulations may either promote or limit the use of intelligent automation technologies dealing with unexpected events (Coombs et al., 2020).

Despite their differences, both IS consultancy companies and international organizations share a common thread in their management of unexpected events: the need for effective communication about their actions and decisions. Academic researchers emphasize the critical role of formal communication in IS (Tsai & Compeau, 2021; Hällgren & Wilson, 2007; Weick et al., 2005). Moreover, Lepistö (2014) notes a scarcity of research explicitly focusing on the discourse employed during IS implementation. Recognizing a research gap in communication influencing IS during unexpected events, Coulon et al. (2023) underscore the importance of addressing this gap as a future research stream in IS unexpected event research. Similarly, Flynn & Du (2012) identify a gap in considering communication and legitimation during challenging situations.

Acknowledging this gap, our central research question emerges as: what are the differences in communication during unexpected events between IS consultancy companies and international organizations ?

To answer this question, we adopt the legitimation discourse angle for investigation. This theoretical framework allows us to analyze not only the content of communications but also the implicit motives, intentions, and strategies employed to legitimize actions for stakeholders (Van Leeuwen, 2007).

This document aims to contribute to the academic legitimation discourse literature by investigating the behavior of IS consultancy companies and international organizations during unexpected events.

The paper begins with a theoretical background to contextualize our study. Moving forward, we pivot to an analysis phase, employing topic modeling as a powerful tool to examine the communication documents of consultancy companies and international organizations. Subsequently, we engage in a nuanced discussion, to interpret and compare the findings from our analysis. Finally, we have a conclusive section that synthesize our observations, highlights key insights, and underscores the potential future stream of research.

## 2 Theoretical background

This literature review is structured as follow, first we introduce the unexpected events, along with its attributes. Subsequently, we introduce the legitimation discourse framework. Finally, we present a comprehensive literature review that encompasses legitimation discourse within the context of IS consultancy companies and international organizations responding to unexpected events.

### 2.1 Unexpected events:

Coulon et al. (2023) defined unexpected events as occurrences that have already happened and are perceived by organizations as having potentially positive or negative impacts, and whose nature or development was not anticipated. In this study we will focus on the events that are inherently linked to IS.

This definition synthesis several key attributes useful for defining unexpected events in IS, signifying an event that has already occurred, was not anticipated, and had negative or positive consequences.

For the first attribute, an unexpected event is an event that has already occurred. Researches employ various terminologies to indicate the past occurrence of the event, such as "realized" (Steffens et al., 2007), "occurred", "happened" and "emerged" (Tukiainen et al., 2010).

Regarding the second attribute, an unexpected event is one that wasn't anticipated. This lack of anticipation can originate during the initial planning stage and may also be reflected in the risk management plans. Pina e Cunha et al. (2006) note that an event can be unexpected in terms of its type (not envisaged by the organization) and/or form (the organization expected the event type but didn't foresee the specific form it would take).

The third attribute, emphasis the potential for unexpected events to result in either negative or positive effects (Alsakini et al., 2004). Recognizing this duality is crucial, as unexpected events may pose threats to the IS, but also offer opportunities for adaptation and innovation. Such events can disrupt schedules and budgets, and impact the quality of deliverables. Beyond operational disruptions, unexpected events can erode stakeholders' confidence in the organization (Aaltonen et al., 2010; Coulon et al., 2013). Moreover, unexpected events can disrupt organizational dynamics, fostering conflicts and contributing to a toxic work environment. Conversely, these events may serve as catalysts for a positive change in the organization's culture. Such transformations can enhance communication openness and encourage a non-punitive approach to errors (Weick, 1998).

Considering these attributes, numerous examples have emerged of unexpected events that can influence various IS stakeholder. For instance, cyberattacks serves as an example, as elucidated by Bélanger & Crossler (2011), who reported that 85 percent of companies experienced privacy breaches in 2007, with 63 percent encountering multiple breaches. These threats have been underscored by the world economic forum (2020), alongside diverse restrictions impacting working conditions for IS organizations. Moreover, Floetgen et al. (2021) presented the effects of the COVID-19 pandemic on IS, shifting priorities from auxiliary to indispensable functionalities. Similarly, this pandemic has led to substantial transformations in IS (Wessel et al., 2021), prompting shifts in client behavior and requirements (Floetgen et al., 2021). Additionally, Sakurai & Chughtai (2020) discuss the impact of the 2011 tsunami earthquake on the utilization of IS in different Japanese cites, where existing IS played a crucial role in organizing and managing rescue efforts. Lastly, DesJardine et al. (2019) explore the repercussions of the 2008 financial crisis on IS projects priorities across various organizations.

## 2.2 Legitimation discourse :

Legitimation discourse, a focal point in discourse analysis elucidated by Van Leeuwen (2007), delves into the use of language to establish legitimacy across various contexts. This concept plays a pivotal role in understanding power dynamics, social constructions, and the negotiation of meaning within different discourses. Van Leeuwen (2007) identified four key categories in his work :

- Authorization Legitimation:

Authorization legitimation, within discourse analysis, involves the use of language to establish and maintain authority or credibility (Van Leeuwen, 2007). Van Leeuwen (2007) outlined different types of this authorization legitimization, such as expert authority (derived from expertise rather than status) and impersonal authority (emerging from laws, rules and regulations).

This category of legitimation is exemplified in the work of Van Leeuwen & Wodak (1999), explaining the authority of the Australian immigration administration in handling family reunion applications. The legitimacy of their decisions can be justified by invoking the authority of law.

- Moral Evaluation Legitimation:

Moral evaluation legitimation employs communication to justify actions or decisions on moral or ethical grounds (Van Leeuwen, 2007). It entails making moral judgments, evaluating actions or decisions as morally right or wrong, good or bad.

Wodak (2006) provides insight into moral evaluation during political debates on immigration in the 2004 US elections and the European Union debates on the concepts of migrant and asylum-seeker. Ethical and moral perspectives are invoked in these debates to legitimize specific decisions and actions.

- Rationalization Legitimation:

Rationalization legitimation entails using reasoning and logic in discourse to legitimize actions or decisions (Van Leeuwen, 2007). it involves offering logical or rational explanations for actions or decisions to present them as justifiable.

Van Leeuwen (2007) discusses how rationalization is employed in media discourse to justify social inequalities. Furthermore, Guild (2002) work illustrate this legitimation category through organizational discourse, where management might rationalize decisions such as employee layoffs as cost-cutting measures for the long-term sustainability of the company.

- Mythopoesis Legitimation:

Mythopoesis legitimation focuses on the creation and dissemination of myths or narratives to legitimize beliefs or practices (Van Leeuwen, 2007). This category contributes to the construction of identity, values, or ideologies, often used for legitimization.

In this context, Vaara (2014) delves into the influence of media discourse in Finland, revealing how discourses of financial capitalism, humanism, nationalism and Europeanism played a central role in legitimation.

In summary, legitimation discourse emerges as multifaceted concept crucial in shaping narratives, ideologies, and power structures.

## 2.3 Legitimation discourse for IS consultancy companies and international organizations in the context of unexpected events

Within academic literature, researchers have explored and documented instances where authorization, moral evaluation, rationalization and mythopoesis legitimations have been employed by IS consultancy companies and international organizations in the context of unexpected events.

- Legitimation discourse for IS consultancy companies

The term "IS consultancy companies" refers to the companies that specialize in constructing, delivering, implementing technology, transferring IS capabilities to the stakeholders, and providing recommendations for optimal business and technical solutions (Winston, 2002).

This literature sheds light on the application of the four categories of legitimation discourse within the realm of IS consultancy companies.

In the domain of authorization, IS consultancy companies employ communications to establish and fortify their legitimacy by emphasizing authority or permission. A noteworthy example is presented by Theiss et al., (2013), illustrating how IS consultancy companies leverage their knowledge of best practices and experience to legitimize their authority in the IS implementation within the complex economic market context. To further support this logic, the researchers clarified that IS consultancy companies consistently justify their authority by highlighting the cumulated knowledge, expertise, skills, and experience of their individual consultants facing unexpected events. Additionally, Hussain & Cornelius, (2009) underline the authority that IS consultancy companies legitimize through the permissions granted by stakeholders on the client side, such as the IS department. This access to strategic organizational information is crucial for keeping the IS aligned with strategic orientations during unpredictable circumstances.

Furthermore, IS consultancy companies utilize the moral evaluation category of legitimation discourse to make moral judgments during unexpected events, assessing actions or decisions as morally right or wrong. Backlund & Werr (2008) provide evidence that, despite potential anxiety and skepticism within the client organization during complex situations, IS consultancy companies commit to their moral judgments and ethics, even when confronted with challenges such as refusal of cooperation, information hiding, and expressions of cynicism. Additionally, Paré et al.,(2020) present the legitimation discourse employed by IS consultancy companies, emphasizing the judgment of the psychological well-being of populations through open communication during unexpected events.

In the rationalization category of legitimation discourse, IS consultancy companies provide logical or rational explanations for actions or decisions. Paré et al.,(2020) highlighted the discourse of IS consultancy companies in health IS tools, presenting results and indicators about the system capabilities and alignment with best practices. Additionally, Hussain & Cornelius (2009) explain how IS consultancy companies communicate to legitimize the use of a new IS tool by justifying its benefits and engaging in discussions to resolve potential conflicts and unexpected situations, emphasizing pragmatism and focusing on solutions that work in the real world within the context of unexpected events. Moreover, Berente et al. (2022) emphasized the legitimacy of IS consultancy companies by communicating the respect the delivery constraints, using the best technology solutions, and effectively managing cost efficiency challenges.

Finally, within the mythopoesis category of legitimation discourse, IS consultancy companies engage in creating myths or narratives that contribute to the construction of identity, values, or ideologies for legitimation. Backlund & Werr (2009) exemplify this by explaining the communication strategies of Accenture and KPMG, promoting an identity of innovation and thinking outside the box during unexpected events. Accenture's website, with the slogan "Accenture–Innovation delivered", emphasizes how their services assist clients in creating their future within a rapidly changing and complex market environment.

- Legitimation discourse for international organizations

The term "international organizations" encompasses a diverse range of formal structures. As defined by Amirci & Cepiku (2020), this term refers to "*all forms of non-state actors working at international or global levels,*" encompassing "*multinational corporations, bilateral organizations, multilateral organizations, and international non-governmental organizations*".

The literature underscores the applicability of the four categories of legitimation discourse in the context of international organizations.

In term of authorization, international organizations employ communication to establish or reinforce legitimacy by emphasizing authority. Discussing the recommendations of international organizations concerning healthcare administration and educational institutions, Flynn & Puarungroj (2006) elaborate on the legitimation process, where authority is used to justify the choice of specific IS tools. The healthcare institution, aligning with the recommendations of international organizations, leverages the authority and permission of a committee composed of department heads. This committee participates in the software selection process to enhance the healthcare system's preparedness of unexpected events. Similarly, Vasconcelos (2007) outlined the authority of university institutions, derived from international organizations' programs and agreements, to implement IS tools for student administration and academic computing services. Legitimizing this authority involves obtaining permission from the operating core, which includes academic departments, support services, and the central office responsible for students' administrative tasks.

Moreover, international organizations engage in the moral evaluation category of legitimation discourse to make moral judgments during unexpected events. Bhat (2013) discusses the moral evaluation of a new IS tool for the Indian central government's unique identification number project, comparing it with international organizations' experiences. The communication addresses moral judgments regarding as data privacy, and technology reliability of this IS. In response to these evaluations, the Indian government establish a biometrics center to ensure data privacy and prevent unexpected accident. Flynn & Hussain (2004) also explain the moral evaluation of a new IS tool in a healthcare organization based on norms from international organizations, especially for the patient's data privacy facing any unexpected events.

Furthermore, the rationalization category of legitimation discourse involves providing logical or rational explanations for actions or decisions. Hsu et al. (2015) provide rational explanations used by multi-cultural interorganizational banks to legitimize their choice of an IS tool. The communication emphasizes that the chosen tool meets international organizations' requirements and reduces the risks of errors during complex situations. Flynn & Hussain (2004) elaborate on international organizations' communication about the importance of new health IS, rising awareness about the benefits and potential uses. Additionally, Vasconcelos (2007) also explains that changing IS tool can promote the standardization of administrative information processes responding to international organizations requirements and improve control over system operations, enhancing the educational institution's efficiency during unexpected events.

Finally, within the mythopoesis category of legitimation discourse, international organizations create narratives to contribute to the construction of identity, values, or ideologies for legitimation. Hsu et al. (2015) explain the cultural narrative promoting the adoption of technology in multi-cultural interorganizational banking information system respecting, the international organizations norms. Similarly, Belanger & Carter (2012) explain how international organizations adopt the narrative of providing citizens with more meaningful services through the adoption of IS tools in the e-government process.

## 3 Methodology

In the preceding paragraph, we provided a literature review on the legitimization of IS consultancy companies and international organizations during unexpected events. This step is essential for understanding the existing research in this area. The next step involves examining the communication documents of IS consultancy companies and international organizations during unexpected events from a legitimization perspective to examine the actual practices in this domain. To achieve this objective, we outline the adopted methodology in this paragraph, utilizing the topic modeling research approach (Blei et al., 2003).

This approach leverages statistical associations among words in a text to generate related topic and clusters. It is very useful in understanding and detecting cultural dynamics, aiding our analysis of diverse communications within our study (Blei & Lafferty, 2007).

For the execution of the topic modeling approach in this study, we utilized the orange data mining software with the version 3.35.0 (Demšar et al., 2013). Our approach involved the following steps : initially, we identified the corpus of materials for consultancies and international organizations. Subsequently, we engaged in corpus analysis using the topic modeling approach. Finally, we analyzed and interpreted the results of our investigation.

### 3.1 Identification of the material corpus :

To explore communication, we selected documents released by IS consultancy companies and international organizations. This involved a comprehensive analysis of the respective websites to identify communication materials, employing these key words together : "information system", "unexpected", "event", "disruptions", "deviations", "changes", and "surprises". Our focus was on recent publications, ranging from 2020 to 2023. We utilized the search features on English language websites, employing the specified keywords. In cases where such a search was unavailable, we manually collected materials by navigating through the website structure. The breakdown of our selection approach is as follows:

- For the IS consultancy companies :

To select the material corpus, the initial step involved identifying IS consultancy companies actively communicating their processes and outcomes in response to unexpected events. Our approach commenced with exploring eligible companies through the vault.com rankings, yielding a preliminary list of 50 potential IS consulting companies. Subsequently, we evaluated documents related to all the listed IS consultancy companies using the criteria explained in the previous paragraph. Through this filtration, we identified 324 materials, including marketing and sales documents, reports, and interviews from 13 IS consultancy companies (Bain, BCG, Deloitte, EY, Kearney, Accenture, KPMG, Mackinsey, Olivier Wyman, PWC, Roland Berger, Gartner). Annex A provides a descriptive study of our selected communication documents of the IS consultancy companies. These documents encompass a total of 1373858 words.

- For the international organizations :

To build the material corpus, the initial step involved identifying a selection of international organizations that communicate their processes and results during unexpected events. Our approach began with exploring eligible international organizations through the list presented by the union of international associations (UIA). As explained by Amirci and Cepiku (2020) in their document, the number of the international organizations, after a filtration process based on the types of organizations, eliminating inactive and dissolved organizations, multilateral treaties, intergovernmental agreements, and organizations emanating from places, persons, and bodies is 285. Subsequently, we evaluated documents related to the listed international organizations using the criteria explained in the previous paragraph.

As a result, our search yielded a total of 205 materials, including thematic articles, white papers, reports, and interviews from 19 international institutions (United Nations, Council of Europe, United Nations Educational Scientific and Cultural Organization (UNESCO), United

Nations International Children's Emergency Fund (UNICEF), World Health Organizations (WHO), Work Economic Form, World Trade Organization, World Bank, European Commission, Asia and Pacific council, North Atlantic Treaty Organization (NATO), European Union Agency for Cybersecurity, Food and Agriculture organization of the united nations (FAO), International Atomic Energy Agency, Arab Monetary Fund, African Development Bank, Asian Development Bank, European Investment Bank, organization for economic co-operation and development (OECD)).

Annex A provides a descriptive study of our selected communication documents relevant to international organizations. These documents englobe a total of 1599871 words.

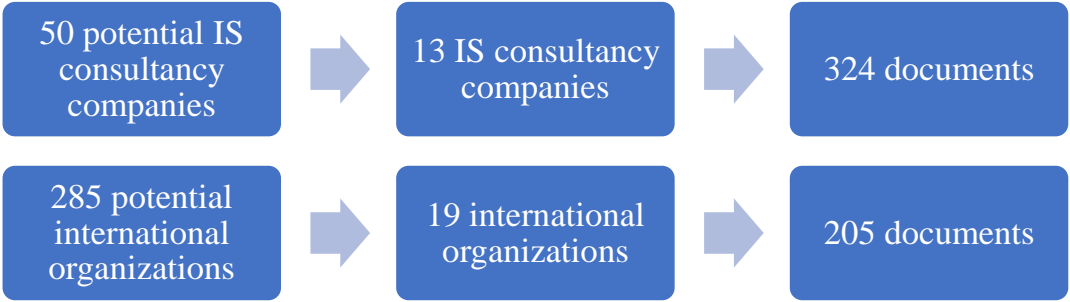

**Figure 1 selection process for the international organizations and the IS consultancy companies' communication documents**

## 3.2  Topic modeling study :

To initiate the topic modeling analysis, we began with a text pre-processing step. Following the approach established by Antons et al. (2016), we meticulously pre-process our text corpus for analysis. References and citations were removed to prevent potential bias in the outcome of our topic model. Employing well-established text mining procedures for topic models (Blei et al., 2003), we further removed stop words, URLs, parsed HTML, punctuation, and numbers, while converting all words to lowercase.

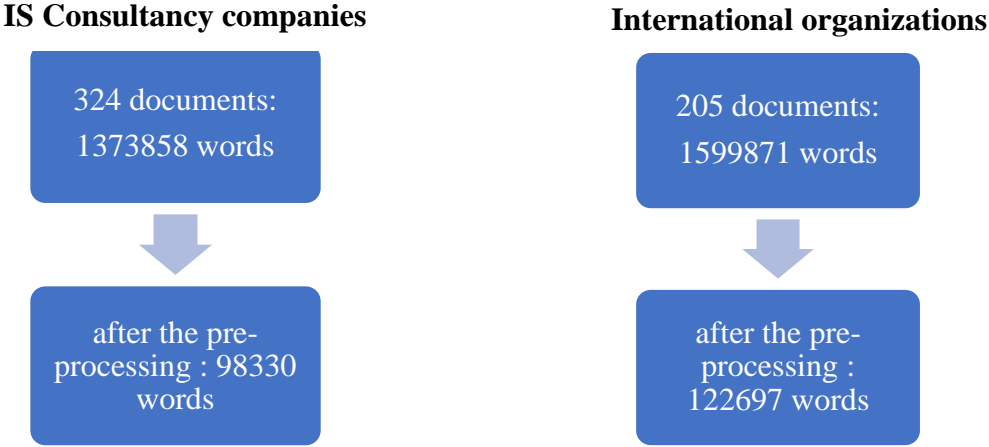

**Figure 2 : Pre - Processing step of analysis**

For the configuration of the topic modeling process, we employed the "latent Dirichlet Allocation (LDA)" technique. This methodology assumes that each document is a composition of various topics, and each topic is a mix of different words (Blei et al., 2003). Given our aim of topic discovery and categorization, this algorithm proves suitable for our topic modeling analysis.

Within the orange data mining software, we designed two distinct workflows (one for the IS consultancy companies' documents and another for international organizations documents).

We initially configured the topic modeling by setting the number of topics to 50, aligned with the recommendation of Blei et al. (2003), which suggests that this parameter grows linearly with the corpus size. Based on the outcomes, we assessed whether the saturation of topics was achieved, subsequently eliminating any duplicated topics. This process yielded 41 topics for IS consultancy companies and 39 topics for international organizations. These topics where further grouped into separate clusters, using the T-SNE diagram for visualization data in the form of a scatter plot. This process yielded six clusters for IS consultancy companies and five clusters international organizations' topics, each encapsulating the main idea of related topics.

Subsequently, we conducted a qualitative exploration of each cluster, focusing on the most relevant document determined by coherence scores and identified keywords related to each legitimation category.

In the next section of the document, we will present the outcomes of this methodology.

## 4 Findings

This section delves into the contextualization of unexpected events for both IS consultancy companies and international organizations. Following our methodological approach, we identified multiple topics relevant to both entities. Subsequently, we presented and analyzed these distinct topics and their clusters.

### 4.1 The context of the study

As a part of the contextualization process outlined by Van Leeuwen (2007), this section aims to illuminate the context of unexpected events for both IS stakeholders: IS consultancy companies and international organizations.

Over the last century, the development of IS organizations has led to an increase in various vulnerabilities (Brumă, 2020). These vulnerabilities, as categorized by Popescu, (2018), encompass concerns related to hardware, software, network, management, personnel and physical (location) infrastructure. In recent years, the frequency of unexpected events impacting IS organizations has escalated, making these vulnerabilities even more threatening. As highlighted in the literature review section, numerous examples of unexpected events with potential impacts on IS organizations has emerged. To effectively manage these unexpected events, there is a pressing need to explore the communication dynamics among IS stakeholders. In this context, our focus is on discerning differences in communication from discourse legitimation perspective.

For IS consultancy companies, these changes manifested in reduced client budgets and a shift in focus towards the sustainability of the existing IS. In response, IS consultancy companies needed to proactively communicate their strategies, success stories, and reassure the market of their capacities to adeptly navigate these unexpected events. Similarly, international organizations found themselves compelled to communicate their strategies to regulating the IS, financing IS companies, and analyzing the impact of unexpected events.

In the subsequent stage of our analysis, we delve into the discourse strategies employed by the selected documents. Across the spectrum of documents of both IS consultancy companies and international organizations, a variety of positions emerged. Some documents showcased IS organizations promoting their success stories during unexpected events, while other critically analyzing these events. Additionally, certain documents explained future actions and plans to address similar events. In the following table, we present the percentage of each position in our document population.

**Table 1 : the main subject of the different documents**

| The IS stakeholder | Success stories | Analysis of the unexpected events. | Perspective for the future. |
| --- | --- | --- | --- |
| IS consultancy companies | 20% | 37% | 43% |
| International organizations | 2% | 43% | 45% |

This corpus of documents encompasses various sources of information that employ one of these discourse strategies and utilize legitimation to justify their positions and actions. Among IS consultancy companies, their documents draw on insights from external IS experts (15%), client feedback to validate the success of their projects (12%), and internal team assessments (73%). Conversely, international organizations leverage external IS experts from different backgrounds, including practitioners and researchers (3%), as well as some external partner organizations and government reports (42%). Additionally, they rely on feedback from their internal analysis teams (55%).

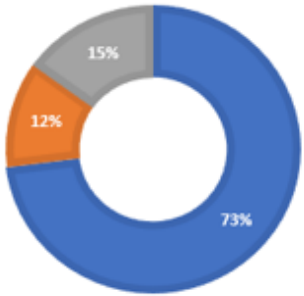
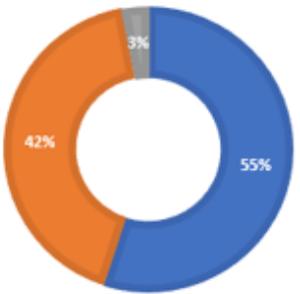

**Figure 3 : Sources of the data in the different documents**

In the next paragraph of the result section, we will delve into the diverse discourse legitimation categories employed in these documents by both IS stakeholders.

## 4.2 The IS consultancy companies

Through the analysis of 324 communication documents, a total of 41 topics were identified, which are detailed in table 1. For an enhanced visual representation, Annex B provide both a word cloud and a T-distributed stochastic neighbor embedding (T-SNE) diagram.

**Table 2 : The list of topics for the IS consultancy companies' communication documents**

| Topic number | Topic name | Topic number | Topic name |
|---|---|---|---|
| 1 | Cultivating the awareness | 22 | Strategies to avoid attacks |
| 2 | Assess ambitions and agility | 23 | Advancing awareness |
| 3 | Agile asset management | 24 | Attractive axes and abilities. |
| 4 | 5G adoption and asset dynamics | 25 | Agile support : aiming to attract through diverse options. |
| 5 | Asset awareness | 26 | Ambitious upcoming years for the agile agents. |
| 6 | Exploring yearly realizations | 27 | The ability to avoid challenges. |
| 7 | Agility in the American market. | 28 | Ensuring technology success |
| 8 | Asia's agencies and assets. | 29 | Evaluating agency and agent. |
| 9 | Improving agriculture : yearly efforts and assurance | 30 | Meeting expectations |
| 10 | Enhancing Aspects and attributes of agility | 31 | Agile approaches : anticipating answers. |
| 11 | Exploring collective goals. | 32 | Agile assessment. |
| 12 | Agility : comparing American and Asian trends. | 33 | Evaluating the ability to accept assistance. |
| 13 | Exploring 3D possibilities | 34 | Building ambitions : agile actions. |
| 14 | Yearly perspective and anticipation | 35 | Agencies agreements overview. |
| 15 | Asian and American contrasts. | 36 | Navigate ambiguity and provide answers. |
| 16 | Yearly asset management | 37 | Exploring Annual Agile Trends. |
| 17 | Anticipation approach | 38 | Increasing awareness about cyber-attacks and security. |
| 18 | Ensuring feedback with an agile approach. | 39 | Shaping future agenda. |
| 19 | Assurance and services overview | 40 | Raise awareness about the management. |
| 20 | Navigating awareness, agility and amplified abilities. | 41 | Asset attacks: awareness, age, and agencies. |
| 21 | Amazon's advancements | | |

From these different topics, we can fetch six clusters as explained in the table 2.

**Table 3 : The list of clusters related to the IS consultancy companies' topics**

| Cluster number | Cluster name | The list of topics related to the cluster. | List of top cluster words |
|---|---|---|---|
| 1 | Cultivate the awareness among | 1, 5, 20, 23, 40, 27, 38 | annual, agencies, aggregate, awareness, associated, avoid, amplified, agree, ambition, agenda, angles, anything, |
| | organizations | | assessment, aware, yield, agreement, |
| 2 | Asset management in time of unexpected events. | 3, 8, 16, 41 | able, year, assets, amazon, ask, asset, assumptions, agency, ahead, avoid, announced, assurance, attention, assistance, attacks, attack, age, attackers, agencies, |
| 3 | Management methods (such as agility) | 2, 7, 10, 12, 18, 25, 31, 32, 37 | assess, ambition, associated, able, assessing, avoid, agility, aerospace, amelioration, attribution, aspects, ahead, anticipate, agile, work, assortment, aspiration, attract, assist, opinion, assigned, annual, asked, |
| 4 | Adaptation of technology | 4, 13, 22, 28, | year, 5G, agile, able, assets, agrees, 3D, avoid, age, attacks, ensure, anomalies |
| 5 | Analyzing the results of the organizations. | 6, 15, 19, 21, 29, 30, 33, 36 | average, year, assessment, associated, able, agenda, amount, aggregation, amazing, assurance, advisory, services, attachments, attended, annual, announced, agency, agricultural, yields, assures, absorb, associations, assistance, agility, attribute, cycle, ambiguous, answer |
| 6 | Future planning and forecasting | 9, 11, 14, 17, 24, 26, 34, 35, 39 | assurance, amelioration, aggregate, assistants, ambition, associated, avoid, agile, aerospace, associated, ask, anticipate, assess, aggressive, assortment, away, able, attractive, agreements, 21st, agent, ambitious, assumptions. |

If we group these clusters, our study reveals the use of three legitimation discourse categories in the communications of IS consultancy companies.

- Moral evaluation legitimation

Among the crucial roles of IS consultancy companies is advising their clients and cultivating awareness about the ethical implications and impacts of technology and management. In the first Cluster, we encompass various topics related to the activities of awareness undertaken by IS consultancy companies, utilizing diverse terminologies such as "awareness", "aware". This proactive engagement enables both IS consultancy companies and clients to navigate unexpected events with a heightened sense of awareness.

For instance, Accenture's documents (2021) indicate a prioritization of data privacy, ethical design, and continuous governance, emphasizing the importance of trust in the moral aspects of IS for the next generation of products and services. McKinsey (2020) stresses the significance of ethics and reflects on how technology can have both positive and negative influences. *"This perspective extends to considerations of data privacy and information protection"*, promoting collective reflection among IS stakeholders.

Additionally, PwC (2022) highlights that organizations *"will be seeking sustainable, ethical digital health solutions and infrastructure, considering social responsibility through the entire supply chain"*. PwC's Digital Health Trends 2022, shaped by a commitment to equity in healthcare, *"reflects a growing focus on sustainability and ethical decision-making"* in response to evolving drivers in the industry.

- Rationalization legitimation

Another crucial aspect is the significance of rationalization legitimation for the IS consultancy companies. Within our study, four clusters were identified : asset management in time unexpected events, management methods, adaptation of technology, and analyzing the results of the organizations. These topics are articulated through varied terminologies, including "evaluate", "agility", "asset", and "amelioration".

Our findings underscore the pivotal role of asset management by IS consultancy companies in navigating unexpected events. These companies play a crucial advisory role, guiding their clients on leveraging available resources effectively during challenging times.

IS consultancy companies communicate their analysis of organizational results, particularly in the context of COVID-19 pandemic. Agile management methods feature prominently in their communications, emphasizing the importance of decision-makers staying responsive to the consequences of unexpected events and adapting decisions accordingly. This methodology fosters flexibility and enables adaptability to changes.

Furthermore, IS consultancy companies actively promote new technologies, highlighting their capacity to implement these technologies rapidly. The feedback from clients already utilizing these technologies reinforces the credibility and accuracy of insights provided.

In PwC's documents (2020), immediate actions for software leaders to optimize performance and preserve shareholder value during unexpected events such as the adoption of agile management methods, are highlighted. *"The software industry is positioned for a rapid recovery from the COVID-19 crisis, with certain segments benefiting from increased adoption of digital solutions to support flexible working, social distancing, and automation"*. PwC (2021) also shares success stories for the implementation of SAP software in client platforms.

Moreover, McKinsey (2020) delves into the challenges faced by public health care IS during the COVID-19 pandemic. They elaborated on the technological and managerial efforts undertaken to overcome these challenges and share feedback on the outcomes of these efforts.

- Mythopoesis legitimation

Mythopoesis legitimation holds significant importance for IS consultancy companies. Within our sixth cluster, these companies have invested substantial effort in crafting forecasts, narratives, and conducting studies on the future of the IS. This effort is reflected in the language used in this cluster, including terms such as "ambitions", "assumptions". This approach not only strengthens the identity and values of the clients but also aids in preparing for forthcoming changes.

In the documents from Accenture (2020), a clear demonstration of their commitment to *"the planet and society is highlighted as an integral part of the company's DNA"*. This identity serves as a narrative, imparting values to both the company and its clients who engage with the companies' services.

Furthermore, Deloitte (2020) showcases the capabilities of new technology in enabling organizations *"across various industries to translate quantitative data into practical business*

*insights"*, constructing narratives and values in the process. Additionally, PwC (2021) presents its cloud transformation solution, emphasizing its alignment with companies' goals and capabilities, thereby contributing to an identity with marketing value for client as well.

### 4.3 The international organizations

Through the analysis of 205 communication documents, a total of 39 topics were identified, which are detailed in table 3. Additionally, in Annex C, we provide both a word cloud and a T-distributed stochastic neighbor embedding (T-SNE) diagram to gain insights into the relationships within our topics.

**Table 4 : The list of topics for the international organizations' communication documents**

| Topic number | Topic name | Topic number | Topic name |
|---|---|---|---|
| 1 | Analysis of the availability of the application | 21 | Application approval and adoption: additional assistance and funding. |
| 2 | Address the availability of assets. | 22 | Awareness attitudes. |
| 3 | Analysis of the applications from an agency approach. | 23 | Advancing associations : awards, adjustments and improvement. |
| 4 | Administration availability and adoption of technology. | 24 | Announcement for the new approaches in the next years. |
| 5 | Approaches to address awareness about technology. | 25 | Ensuring applications against attacks. |
| 6 | Required funding for the announced development agendas. | 26 | Agreements for the assistance and availability of the systems. |
| 7 | Technology adoption agendas | 27 | Analysis for Asian assets. |
| 8 | Administration approaches in addressing expansion and availability among agencies. | 28 | Avatar analysis |
| 9 | Agreements and average insights | 29 | Analysis for the available approaches in Asia. |
| 10 | Evaluating the announcements. | 30 | Adapting agro ecosystems. |
| 11 | Approaching attack assessment | 31 | Application analysis: addressing monetary aspects. |
| 12 | Applications and assets : insights for users | 32 | Raise awareness about cyberattacks. |
| 13 | Advancing agriculture : agrotech adoption | 33 | African administration : adopting American approaches. |
| 14 | Addressing appropriate agencies and administration | 34 | Navigating future agendas : approaches, awareness, and assurance. |
| 15 | Application announcement for the tech giant's (amazon and apple). | 35 | API technology adoption. |
| 16 | Administration strategy. | 36 | Availability for the US administration. |

| 17 | Future work, improvement, adaptation, and associations. | 37 | Answers for the administration review. |
| 18 | Adaptation and engagement of the users. | 38 | Administration's average availability and absence. |
| 19 | Evaluating average users. | 39 | The evolution of systems between 20$^{th}$ and 21$^{st}$ century. |
| 20 | Azerbaijan's asset landscape | | |

From these different topics, we can fetch many clusters as explained in the table 4.

**Table 5 : the list of clusters related to the international organization's topics**

| Cluster number | Cluster name | The list of topics related to the cluster. | List of top cluster words |
|---|---|---|---|
| 1 | Analysis and evaluation of the IS results. | 1, 2, 3, 10, 20, 19, 18, 28, 27, 29, 39, 11, 12, 15, 17, 23, 24, 35, 7, 13, 25, 30 | available, analysis, applications, adoption, attack, approach, able, assets, address, associated, additional, agencies, year, agenda, average, 2020, 4G, apple, announce, 5G, attack, appropriate, agriculture, AGRITECH, future, amelioration, adaptation, attended, adjustments, agree, annual, approximately, assess, award, assembly, assurer, analyst, availability, versions, avatars, axis, 3D, amount, applied, appropriate, adaptive, AGRO-ecosystem, adapt, API, 20th, 21st, |
| 2 | Administration approach for IS | 4, 8, 14, 16, 33, 36, 37, 38 | administrator, available, adopted, applications, assessor, addition, administration, administrative, address, approach, average, agency, agreement, appropriate, able, strategy, ambassador, adaptable, attributed, assistance, adapt, agenda, annually, answers, added, absence, |
| 3 | Raise the awareness about IS | 5, 22, 32, 34 | awareness, address, applications, able, aware, attitudes, attacks, approach, agency, safe, privacy, analysis, agencies, appropriate, agenda, aspects, insurance |
| 4 | Funding for the IS | 6, 21, 31 | average, amount, funding, announced, year, approved, application, apply, finance, payment, agreement, additional, assistance, appropriate, analyzing, address, money, trends |
| 5 | Legislations of IS | 9, 26 | agree, adult, attitudes, average, regulation, legal, agreement, available, approach, address, analysis, assistance, assessment |

If we group these clusters, our study reveals the use of three legitimation discourse categories in the communications of IS consultancy companies.

- Authorization legitimation

In our study, the theme of authorization legitimation manifests across three clusters : Administration approach for IS, Funding for the IS, and Legislations of IS. These clusters incorporate a diverse range of terminologies such as "regulation", "funding", "administrative", "approved", and "agreement".

Institutions concentrate on the administrative approach of IS, with a specific emphasis on ensuring the availability of critical systems, such as health and insurance IS, during challenging scenarios like COVID-19 pandemic.

These capacities empower institutions to maintain a global perspective on ongoing projects, facilitating regulatory and financial actions. For example, they enact regulations dictating accounting practices, crucial for establishing an efficient taxation system and necessitating adjustments in company IS. Another instance involves the regulation of tech giants like Amazon and Apple.

Within the documents of the world bank group (2021), *"the global response to the COVID-19 health crisis triggered governmental actions, such as lockdowns and increased reliance on contactless financial products and services"*. *"This accelerated the shift to digital finance in many economies"*, showcasing how governments utilized their authority to regulate and support digital payments development.

Furthermore, in the OECD (2021) document, numerous countries and authorities exerted significant efforts to transition the educational system online during the COVID-19 crisis, resulting in the total digitalization of the educational process.

The analysis of the European parliament (2020) sheds light on the primary legal and regulatory challenges posed by the applications of these technologies in the context of a public-health emergency. *"Covid-19, as the first pandemic of the century, provides an excellent opportunity for policymakers and regulators to reflect on the legal plausibility and effectiveness of deploying emerging technologies under time pressure"*.

- Moral evaluation legitimation

The moral evaluation legitimation is presented in this study by the third cluster addressing the heightened awareness about IS. This cluster use different words such as "awareness", "safe", and "privacy". In this cluster, international organizations evaluate IS from ethical and moral perspectives. Through their international programs, these institutions can assess the IS of numerous countries, spanning both affluent and economically challenged nations. Their capacity extends to analyzing the impact of IS during major unexpected events, including wars (examining the IS of ONG's providing medical aid and basic needs to the refugees), natural disasters (assessing the role of IS in organizing rescues missions), and economic risks (providing an overview of IS in banks and companies).

Within the documents of the OECD (2021), governments and public authorities leverage digital resources to enhance ethical awareness and promote the safe use of IS. This involves initiatives such as education websites, digital financial education, and social media campaigns, tapping into influencers and stars for broader reach.

Furthermore, the United Nation (2022) emphasizes the use of *"emerging technologies, particularly artificial intelligence and 5G technology, in a convergent and interoperable manner. The focus is on considerations related to ethics, impartiality, transparency, accountability, security, privacy and non-discrimination"*.

The World health organization (2021) classifies *"health data as sensitive personal data requiring high safety and security standard"*. It underscores the importance of ethics in ensuring the collection of good-quality data, advocating for transparency and effective communication about data security strategies.

- Rationalization legitimation

The rationalization legitimation is presented in this study through the first cluster. This cluster delves into the analysis and evaluation of IS outcomes, employing varied terminologies such as "analysis", "amelioration", and "adjustment". In this segment, international organizations communicate their actions and decisions, showcasing IS success stories that emphasize the efficiency of technologies. They highlight instances where international organizations interventions rescued struggling IS companies in unexpected events and advocate for investments in IS and technologies, particularly those associated with the agriculture sector (Agritech).

In the documents of the World economic forum (2023), there's a comprehensive presentation on the synergy of *"artificial intelligence and emerging technologies with agriculture. This synergy provides a remarkable opportunity for India to enhance its agriculture by making it more agile, data-driven, and efficient, all while prioritizing farmers' well-being"*, addressing food security, and aligning agriculture with climate challenges.

Furthermore, the World economic forum (2022) report indicates that 81% of cyber leaders believes that the digital transformation encouraged by international organizations is a key driver in enhancing cyber resilience. *"The accelerated pace of digitalization, stimulated by the COVID-19 pandemic and changes in work habits, is propelling cyber resilience forward"*.

## 5 Discussion and conclusion

In the preceding analysis, we analyzed the communications of IS consultancy companies and international organizations through the lens of legitimation discourse as presented by Van Leeuwen (2007). Each stakeholder, in their communications, cited their actions during unexpected events, seeking to justify these actions from their unique perspective. Both stakeholders asserted that their actions effectively guided IS companies toward better management of challenging situations, minimizing losses, and even uncovering opportunities.

Comparing the categories of legitimation discourse used in the literature with the results of our analysis for IS consultancy companies and the international organizations reveals some distinctions. IS consultancy companies incorporate moral, rationalization, and mythopoesis legitimations in their communication documents. This can be attributed to their emphasis on tangible achievements and logical arguments during unexpected events, diverging from extensive focus on permissions and authority.

In contrast, international organizations primarily concentrate their communication on authority, permission, results, and moral orientation, rather than relying on the creation of myths or narratives to legitimize their actions.

In response to the research question, our legitimation discourse perspective unveils distinct communication strategies adopted by IS consultancy companies and international organizations in face of unexpected events. IS consultancy companies, functioning as strategic partners, communicate proactively to clients, offering strategies for leveraging assets and implementing technology. They draw on market experience and insights, addressing IS challenges arising from unexpected events and presenting innovative solutions. This proactive approach positions them as providers of predictive theories and proactive measures for upcoming events, showcasing their practicality, efficiency, and effectiveness.

On the other hand, international organizations, playing a supervisory role, communicate about regulations, fund projects aligned with their planning, and raise awareness about specific technologies. They highlight their role as supervisors, entrusted with guiding IS through international legal frameworks during unexpected events. This represents moral legitimation, as they adopt a broader perspective on projects, ensuring collective benefits. Moreover, by subsidizing IS research or aiding struggling IS markets, they significantly contribute during periods like the COVID-19 pandemic.

This study makes a valuable contribution to the academic research by addressing a gap highlighted by Coulon et al. (2023), Flynn & Du (2012) and Lepistö (2014), emphasizing the need to delve deeper into communication dynamics influencing IS during unexpected events. Our approach involves a comprehensive examination of the communication employed by both IS consultancy companies and international organizations, adopting a legitimation perspective as proposed by Van Leeuwen (2007). Both IS stakeholders presents a unique opportunity to gain insights into communication practices during unexpected events. IS consultancy companies, with their focus on feedback, innovative services, and documented solutions to overcome challenges (Stockhinger & Teubner, 2018), complement the supervisory and evaluative mechanisms employed by international organizations (Guy, 2020 ; Coombs et al., 2020). From a managerial perspective, this study enhances our understanding of the communications mechanisms of these two IS stakeholders. This deeper insight can provide management teams with a better understanding, enabling them to make informed decisions regarding the communications strategies of both stakeholders.

One significant limitation of our study is related to the sampling and its representativeness. Although our aim was to encompass a diverse range of unexpected events influencing IS organizations, we encountered challenges in diversifying the types of events. The study period from 2020 to 2023 predominantly focused on events such as the COVID-19 pandemic and cyberattacks, limiting the variety of scenarios analyzed. Furthermore, our study relies on the analysis of communication documents released by various IS organizations. To enrich our understanding, incorporating interviews could offer valuable insights into the legitimation discourse processes adopted by different IS stakeholders.

To further explore this topic, future research could investigate the interactions between international organizations and IS consultancy companies through their communications. Such a study could enhance our understanding of both stakeholders, exploring potential conflict subjects and collaboration opportunities in the context of unexpected events.

Additionally, examining how these legitimation discourses manifest within organizational settings could enhance our understanding of IS organizational dynamics during such events. Moreover, exploring how the legitimation of IS is perceived by employees who utilize these systems within enterprises would provide valuable insights into its reception and impact.

**Annex**

## Annex A : Descriptive study of the corpus material

**Table 6 : the different documents of the IS consultancy companies**

| The IS consultancy company | Number of documents |
|---|---|
| Bain | 12 |
| BCG | 22 |
| Deloitte | 31 |
| EY | 9 |
| Kearney | 37 |
| Accenture | 21 |
| KPMG | 45 |
| Mackinsey | 48 |
| Olivier Wyman | 31 |
| PWC | 148 |
| Roland Berger | 9 |
| Gartner | 11 |

Table 7 : the different documents of the international organizations

| The international organizations | Number of documents |
|---|---|
| United Nations | 8 |
| Council of Europe | 12 |
| United Nations Educational Scientific and Cultural Organization (UNESCO) | 17 |
| United Nations International Children's Emergency Fund (UNICEF) | 5 |
| World Health Organizations (WHO) | |
| World Economic Forum | 38 |
| World Trade Organization | 8 |
| World Bank | 13 |
| European Commission | 14 |
| Asia and Pacific Council | 5 |
| North Atlantic Treaty Organization (NATO) | 8 |
| European Union Agency for Cybersecurity | 3 |
| Food and Agriculture Organization of The United Nations (FAO) | 4 |
| International Atomic Energy Agency | 7 |
| Arab Monetary Fund | 10 |
| African Development Bank | 8 |
| Asian Development Bank | 12 |
| European Investment Bank | 14 |
| Organization for Economic Cooperation and Development (OECD) | 19 |

## Annex B : Topic modeling results for the IS consultancy companies

Figure 4 : word cloud for the IS consultancy companies' analysis

**Figure 5 : Clusters for the IS consultancy companies' analysis**

**Annex C : Topic modeling results for the international organizations**

**Figure 6 : word cloud for the international organizations' analysis**

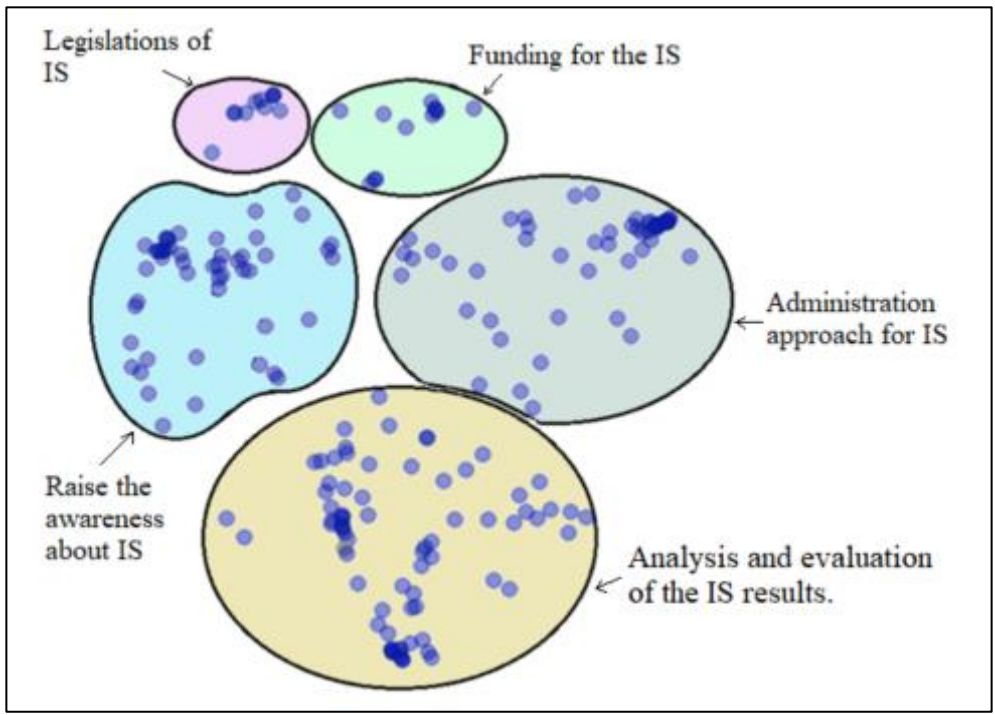

**Figure 7 : Clusters for the international organizations' analysis**